\documentclass[prd, preprint, 12pt]{revtex4-1}

\usepackage{amsmath}
\usepackage{amssymb}
\usepackage{setspace}
\usepackage{graphicx}
\usepackage{bbm}
\usepackage{float}
\usepackage{hyperref}
\usepackage{braket}

\begin{document}
 
 %

\begin{center}
 {  \large {\bf Space-time from Collapse of the Wave-function}}



{\bf Tejinder P. Singh}

{\it Tata Institute of Fundamental Research,}
{\it Homi Bhabha Road, Mumbai 400005, India}

\end{center}
\setstretch{1.24}

\vskip 0.4 in

\centerline{\bf ABSTRACT}
\medskip
\noindent We propose that space-time results from collapse of the wave function of macroscopic objects, in quantum dynamics.
We first argue that there ought to exist a formulation of quantum theory which does not refer to classical time. We then propose such a formulation by invoking an operator Minkowski space-time on the Hilbert space. We suggest relativistic spontaneous localisation as the mechanism for recovering classical space-time from the underlying theory. Quantum interference in time could be one possible signature for operator time, and in fact may have been already observed in the laboratory, on attosecond time scales. A possible prediction of our work seems to be  that interference in time will not be seen for `time slit' separations significantly larger than  100 attosecond, if the ideas of operator time and relativistic  spontaneous localisation are correct.



\noindent 



\bigskip


\section{Introduction}  In a recent paper \cite{Singh:2018} we have proposed that space-time arises as a consequence of localisation of the wave function of macroscopic objects due to the dynamical mechanism of spontaneous localisation. In the present paper we present the same result, along with new physical insights and an experimental prediction, from a different perspective. We start by noting that there is a `problem of time in quantum theory'. One possible resolution of this problem is to invoke an operator space-time in which time is no longer a classical parameter. Classical space-time, along with classical matter, is recovered from operator space-time by invoking a relativistic generalisation of spontaneous collapse of macroscopic objects. In so doing, we predict the new phenomena of spontaneous localisation in time, and quantum interference in time, which should be looked for in laboratory experiments. We explain how the standard quantum theory on a classical curved space-time background is recovered from an underlying quantum theory on an operator space-time, by suppressing the operator nature of time. The originally  proposed resolution of the quantum measurement problem via spontaneous collapse \cite{Ghirardi:86} is seen as an inevitable 
by-product of the relativistic spontaneous localisation that we propose in the present work to recover classical space-time from operator space-time.
\bigskip
\section{The need for a formulation of quantum theory without classical space-time}
Dynamics as we know it  can be very roughly divided into two classes: classical dynamics on a classical space-time, and quantum dynamics on a classical space-time. This is depicted in the cartoon in Fig. 1 below.
\begin{figure}[H]
	\centering
	\includegraphics[width=1.0\linewidth]{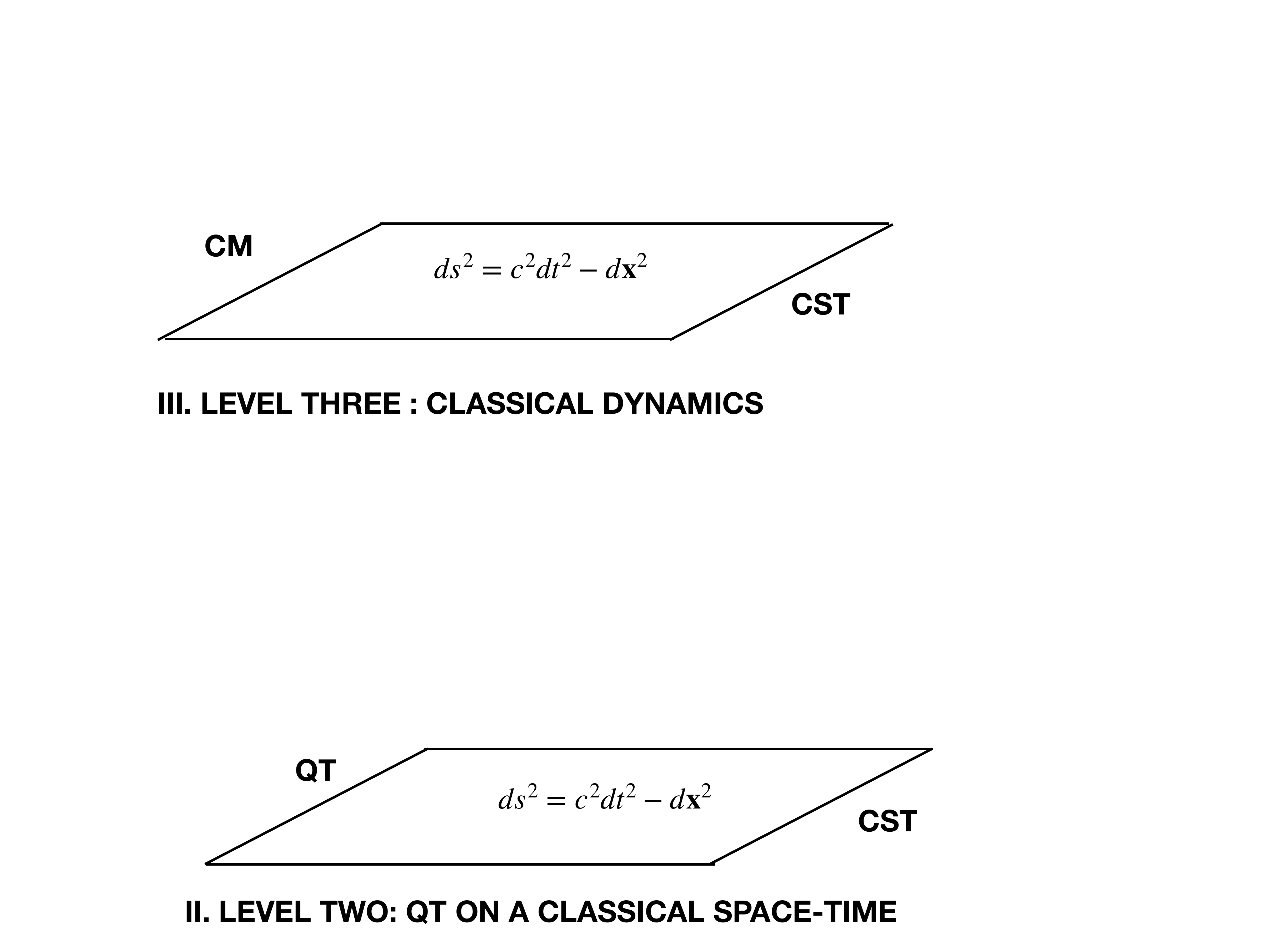}
	\caption{A rough classification of dynamics. Level III. is Classical Mechanics (CM) on a Classical Space-Time (CST).  Level II. is Quantum Theory (QT) on the same classical space-time CST.}
\end{figure}
Level III. in this figure symbolically depicts/includes Newtonian mechanics and Galilean relativity, special relativity, and also general relativity. The curved-space metric is suppressed for simplicity, the key emphasis being that classical objects and fields produce and co-exist with classical space-time.

Level II. in this figure symbolically depicts quantum theory on classical space-time, and includes non-relativistic and relativistic quantum mechanics, quantum field theory, and quantum field theory on a curved space-time. The key emphasis here is the assumption that quantum systems can co-exist with  a classical space-time. At a fundamental level, this assumption is problematic, as is depicted in Fig. 2 below \cite{Singh:2012}.
\begin{figure}[H]
	\centering
	\includegraphics[width=1.0\linewidth]{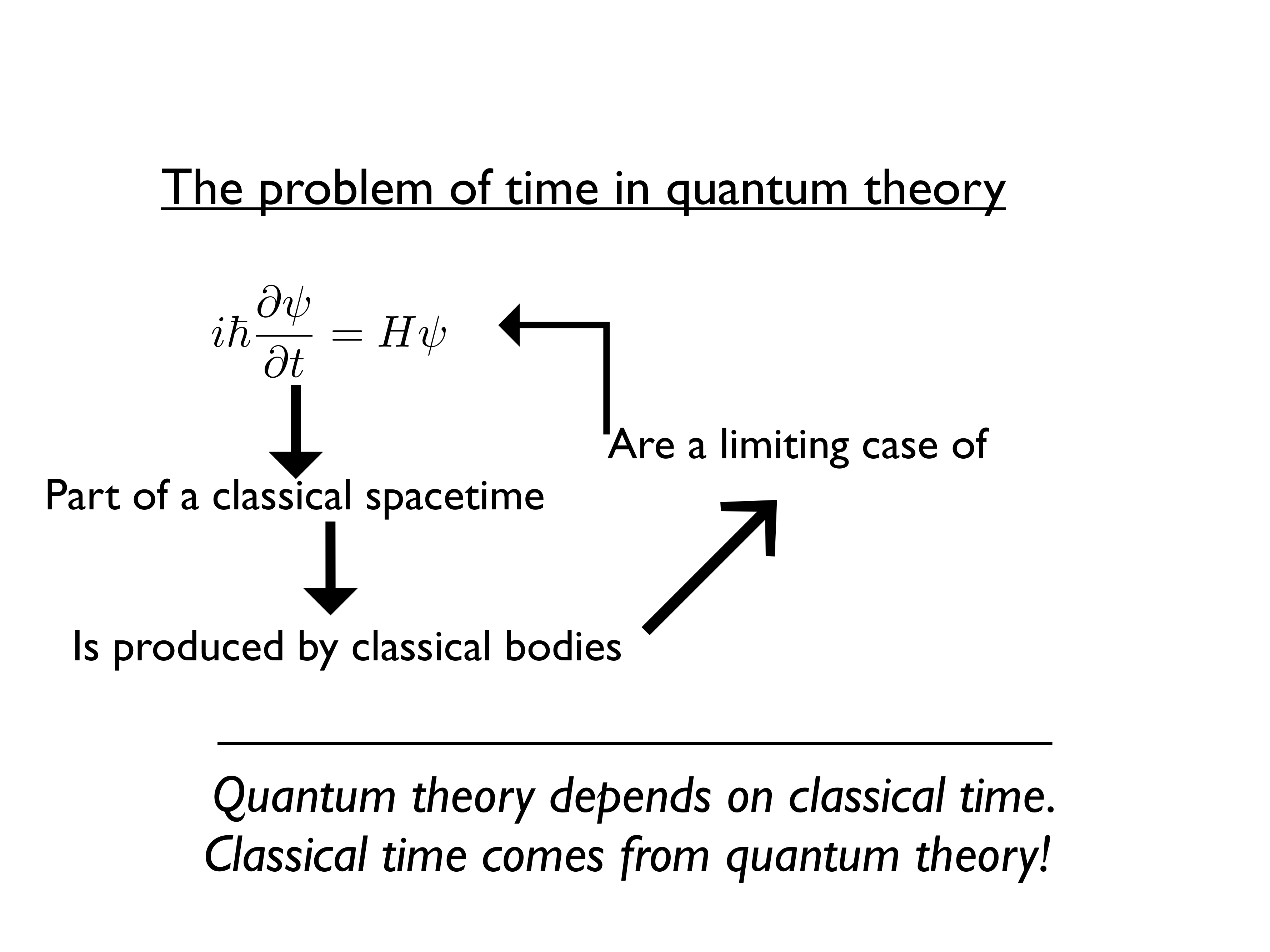}
	\caption{The problem of time in quantum theory.}
\end{figure}
The time parameter which keeps track of evolution in quantum theory, is part of a classical space-time manifold, whose overlying geometry is produced by classical bodies. But these classical bodies are in turn a limiting case of quantum theory, thus making quantum theory depend on its own classical limit. It is a consequence of the application of the Einstein hole argument that if only quantum systems are present, one will have quantum fluctuations in the metric, and as a result one cannot give physical meaning to the point structure of the underlying space-time manifold \cite{Carlip2001}. Thus level II. in Fig. 1 is only an approximate/effective description of the dynamics and it requires the dominant pre-existence of classical matter fields in the universe. At a fundamental level, where there are no classical systems,  there ought to exist a formulation of quantum theory which does not refer to classical space-time. We call this Level I. It follows that Level II. should be arrived at from Level. I, in a suitable approximation.

\section{A possible formulation of quantum theory without classical space-time}
We would like to make a minimal departure from classical space-time, in order to arrive at Level I. Ignoring gravity for the present, we assume that there is a Minkowski space-time metric on Level III. We then propose that physical laws are invariant under inertial coordinate transformations of {\it non-commuting} coordinates, which now acquire the status of operators (equivalently matrices), (${\hat t}, {\hat{\bf{x}}}$). The transition from Level II. to Level I. is made by bringing in non-commutativity of the coordinates, with the coordinates obeying arbitrary commutation relations. There is thus an operator space-time, and from the operator line-element a scalar Trace time $s$ is defined as follows:
\begin{equation}
ds^2 = {\rm Tr}\; d\hat{s}^2 \equiv {\rm Tr} [c^2\;  d\hat{t}^2 - d {\bf \hat {x}^2 } ] 
\label{ost}
\end{equation}
In analogy with  special relativity, one can construct a Poincare-invariant dynamics for the operator matter degrees of freedom which live on this space-time. We call this a non-commutative special relativity - it is a classical matrix dynamics on an operator space-time. Given a Lagrangian for the system, one can write down the equations of motion, where time evolution is now recorded by the Trace time. And one can write down Hamilton's equations of motion for the canonical position operators and their conjugate momenta operators, like in conventional classical mechanics \cite{Lochan-Singh:2011}.

One then constructs a statistical thermodynamics for these matrix degrees of freedom, following the theory of Trace dynamics developed by Adler and collaborators \cite{Adler:04}. Remarkably, it is shown that at thermodynamic equilibrium, the thermal averages of the fundamental degrees of freedom obey the rules of relativistic quantum theory.  But this is now on the operator space-time metric (\ref{ost}), with the operator coordinates now commuting with each other and with  the matter degrees of freedom. Evolution is still recorded by the trace time. Following the techniques of Trace dynamics, one can develop a relativistic quantum field theory on this operator space-time. However, for our present considerations, we will restrict ourselves to a many particle relativistic system. Given a system with $n$ matrix degrees of freedom labelled $q_i^\mu$, it obeys a Lorentz invariant Schr\"{o}dinger equation for the wave-vector $\ket\psi$, which evolves with Trace time, and the index $\mu$ signifies that $q^{\mu}_{i}$ is the `position' four-operator in operator space-time, for the $i$th particle. This quantum dynamics on the operator space-time is our sought after formulation of quantum theory without classical space-time \cite{Lochan:2012}. We could have written this down straight away, but starting from non-commutative special relativity elegantly shows the underlying symmetry, and the minimal departure from classical space-time  that is introduced  by non-commutativity of coordinates. This is the desired Level I, and it is depicted in Fig. 3 below.
\begin{figure}[H]
	\centering
	\includegraphics[width=1.0\linewidth]{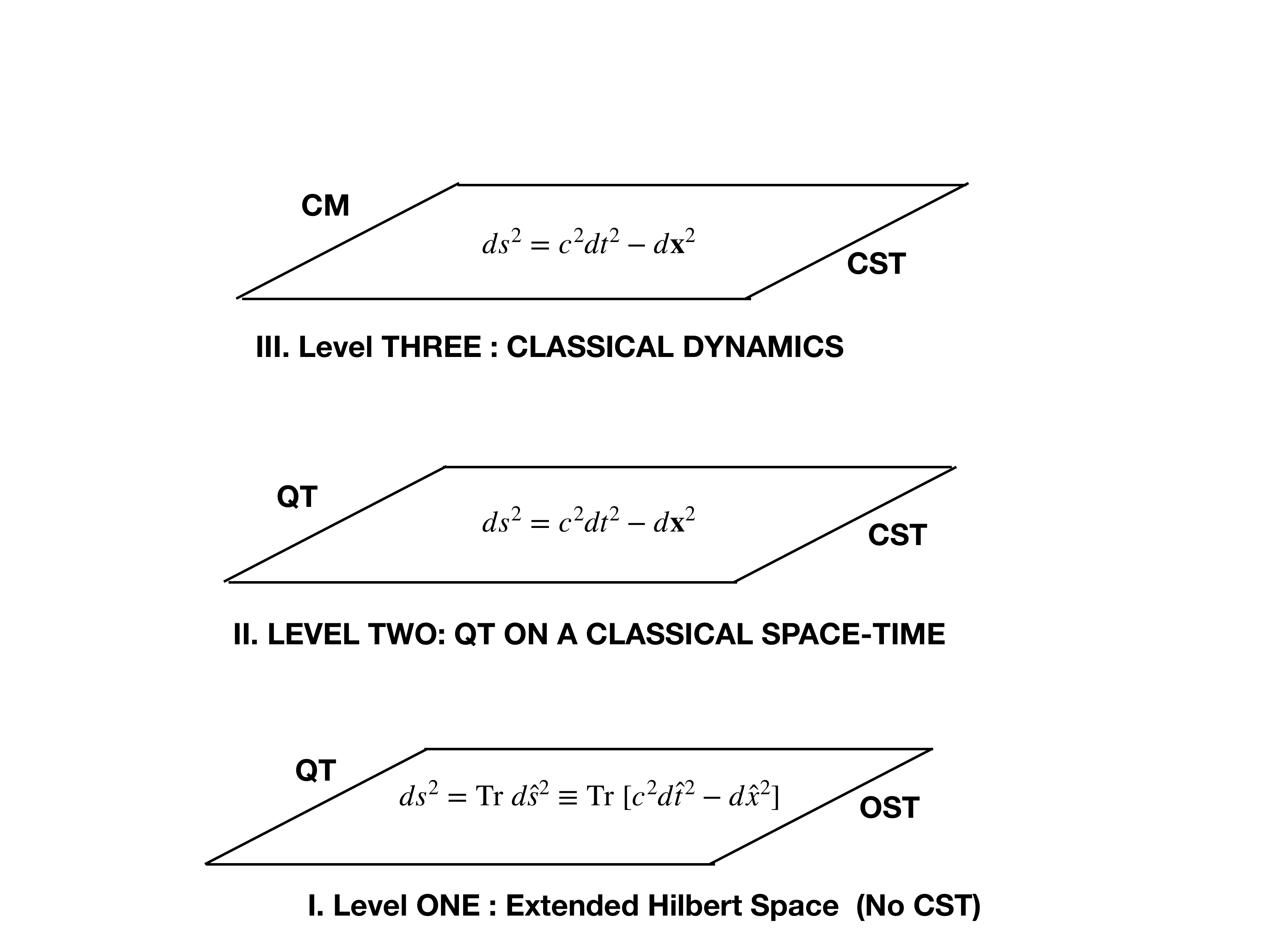}
	\caption{Introducing Level I. Quantum theory without classical space-time, and the extended Hilbert space. Here, classical space-time is replaced by the Operator Space-Time (OST), which transforms the Hilbert space of quantum theory to the Extended Hilbert Space.}
\end{figure}
Level I. has a very significant feature. It is that the Hilbert space, endowed with the operator metric, is now the entire physical universe. There is no more any classical physical  space or classical  space-time, outside this `Extended Hilbert Space'. Thus there is no longer the uneasy tension between the conventional quantum Hilbert space on the one hand - where the wave-function resides - and the particles on the other hand, which this wave-function is supposed to describe, but which live in physical 3-space. In the standard picture, the Hilbert space and the physical 3-space have no apparent physical connection. By endowing the Hilbert space with an operator metric, we overcome that discord \cite{Singh:2018}.

We must now understand how to descend from Level I. to Levels II. and III. First we propose that a transition has to be made from Level I. to Level III. (see Fig. 4 below). This is done by invoking a relativistic generalisation of the spontaneous collapse mechanism of the Ghirardi-Rimini-Weber (GRW) theory,
\begin{figure}[H]
	\centering
	\includegraphics[width=1.0\linewidth]{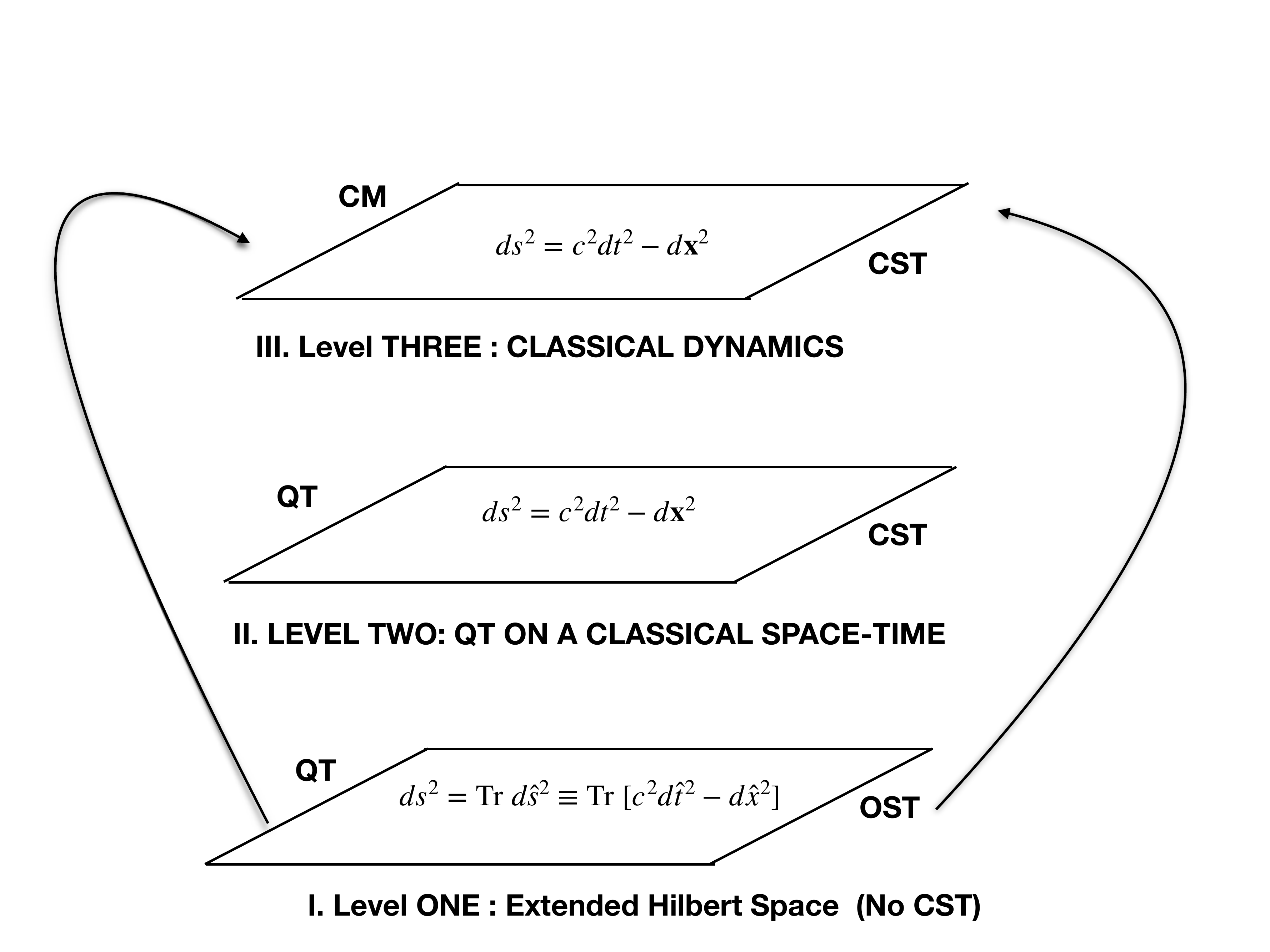}
	\caption{Recovering Level III.  from Level I. by invoking relativistic spontaneous localisation.}
	\end{figure}
\section{Space-time from collapse of the wave-function}
We define the self-adjoint space-time operator $\hat{x}^\mu$ as $x^\mu = (\hat t, {\bf \hat x})$, where all the four operators commute with each other and with the ${\bf \hat q_n}$s. In the `position' representation, the state  of the system is labelled by eigenvalues of $\hat{x}^{\mu}$, and is hence written as
$\psi(x^\mu_1, x^\mu_2, ..., x^\mu_N)$. Evolution is governed by the trace time $s$ defined above.
The dynamics is then given by the following relativistic generalisation of the two GRW postulates \cite{Singh:2018}.

1. Given the wave function $\psi(x^\mu_1, x^\mu_2, ..., x^\mu_N)$ of an $N$ particle quantum system in extended Hilbert space, the $n$-th particle undergoes  a `spontaneous collapse' to a random eigenvalue $x^{\mu}$ of ${ \hat x}^\mu$, as defined by the following jump operation:
\begin{eqnarray}
{\psi_{s}(x^\mu_1, x^\mu_2, ..., x^\mu_N)\quad
\longrightarrow \quad} 
 \frac{L_{n}({x^\mu}) \psi_{s}(x^\mu_1, x^\mu_2, ..., x^\mu_N)}{\|L_{n}({x^\mu}) \psi_{s}(x^\mu_1, x^\mu_2, ..., x^\mu_N)\|}
\end{eqnarray}

The jump operator $L_{n}({x^\mu})$ is a Lorentz invariant linear operator defined to be the normalised Gaussian:
\begin{equation}
L_{n}(x^\mu) =
\frac{1}{(\pi c t_C)^{2}} e^{- ({
{ \hat q}^\mu_n} - {x^\mu})^2/2 c^2 t _C^2}
\end{equation}
$\hat{q}^\mu_{n}$ is the position operator for the $n$-th particle of the system and the random variable ${x^\mu}$ is the eigenvalue of ${\hat x}^\mu$  to which the jump occurs. $t_C$ is a new constant of nature.

The probability density for the $n$-th particle to jump to the eigenvalue  ${x^\mu}$ of ${ \hat x}^\mu$ is assumed to be given by:
\begin{equation}
p_{n}({ x^\mu}) \quad \equiv \quad \|L_{n}({x^\mu}) \psi_{s}(x^\mu_1, x^\mu_2, ..., x^\mu_N)\|^2
\end{equation}
Also, it is assumed  that the jumps are distributed in trace time $s$ as
a Poissonian process with frequency $\eta_{GRW}$, which is the second new constant of the model.

2. Between two consecutive jumps, the state vector evolves according to the following generalised Schr\"odinger equation
\begin{equation}
i\hbar \frac{\partial\psi}{\partial s} = H \psi (s)
\label{ose}
\end{equation}
The particles described by ${q^\mu_n}$ `live' in the ${ \hat x}^\mu$ operator space-time, and the aforesaid ${x^\mu}$ values are actually eigenvalues of ${\hat{x}^\mu}$. Rapid collapse localises a macroscopic object  to one of the eigenvalues of ${ \hat{x}^\mu}$. Using these eigenvalues as reference points, one interprets the collection of eigenvalues as the four dimensional classical space-time we are  familiar with. Space-time could be said to be that which is between GRW jumps in the operator space-time. A quantum mechanical particle which has not undergone collapse still `lives' in the space-time operator space ${ \hat{x}^\mu}$. 
Classical space and time are thus approximations to the operator space and time described by $({\bf \hat{x}}, \hat{t})$, the approximation being caused by GRW quantum jumps.  One can consider the classical  line-element $(c^2 dt^2 - d{\bf x}^2)$
to be one of the eigenvalues of the operator line element $( c^2\;  d\hat{t}^2 - d {\bf \hat {x}^2 } )$ and the Lorentz invariance of the latter ensures the Lorentz invariance of the former. The proper time of special relativity can be said to be the classical correspondence of Trace time. The transition from Level I. to Level III. is depicted in Fig. 5 below. In the process, Level II. is bypassed - we return to Level II. in the next section. It is evident from this figure that we actually live in the Extended Hilbert Space
\begin{figure}[H]
	\centering
	\includegraphics[width=1.0\linewidth]{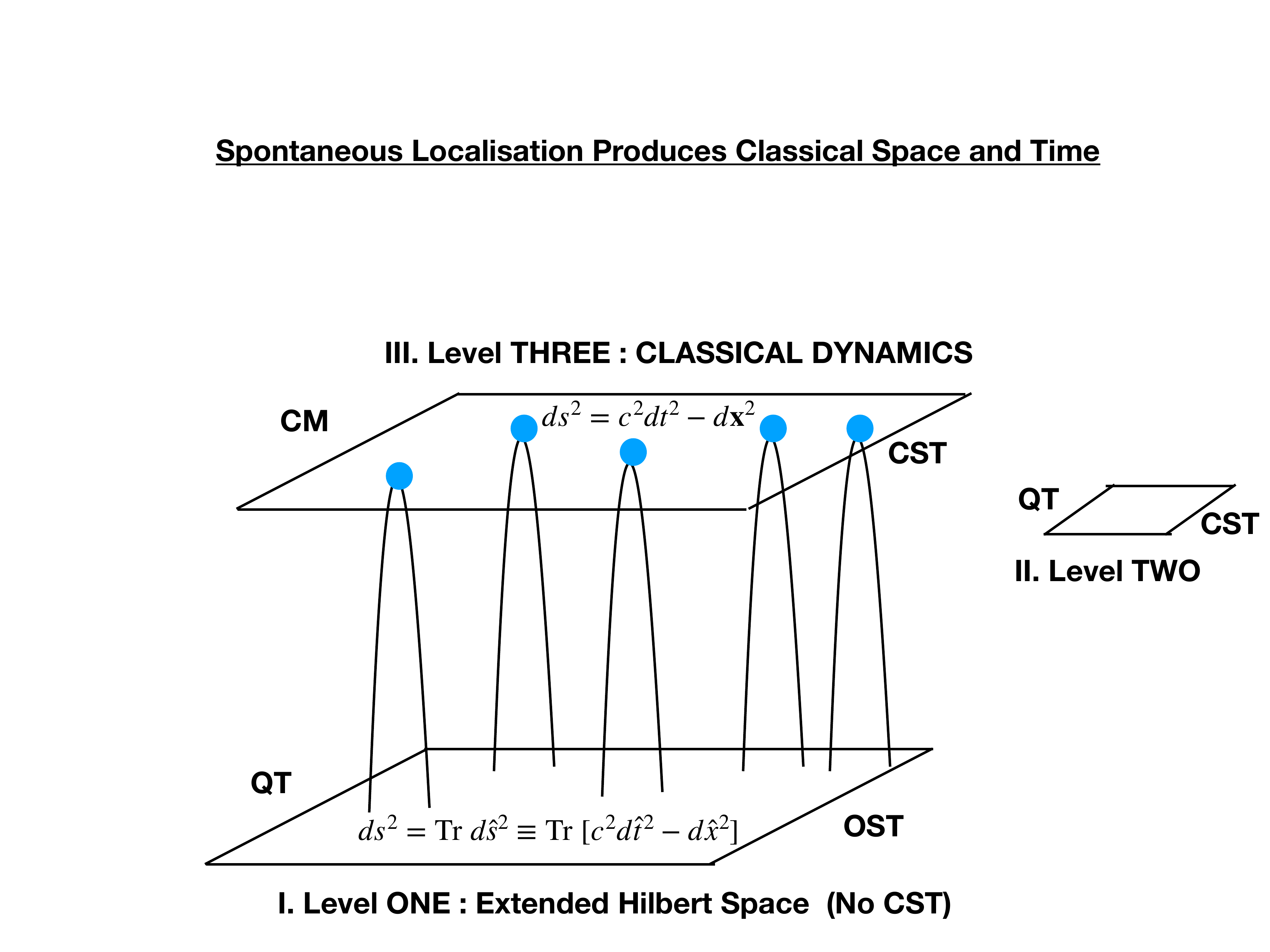}
	\caption{Recovering classical space-time of Level III.  from Level I. by invoking relativistic spontaneous localisation of macroscopic objects.}
	\end{figure}
Just as a macroscopic object spontaneously collapses to a specific position in space and repeated collapses keep it there, spontaneous collapses in time keep it frozen at a specific value of classical time. How then does it evolve in time? This is a serious difficulty with the model as it stands. One possible solution is to propose that spontaneous collapse takes place not onto space-time points, but to space-time paths. Paths are more fundamental than points. Instead of constructing paths from points, we should construct points from paths. Evolution in time is then a perception - the entire space-time path is in fact pre-given, in the spirit of the principle of least action, which determines the entire path in one go. The mathematical formulation of this proposal is presently being attempted.

It is also interesting to note that starting from non-commutative special relativity on level I. one could consider recovering the usual special relativity at Level III., perhaps by a mechanism analogous to spontaneous localisation. This process entirely bypasses quantum theory, and might be worthy of further investigation.

\section{Recovering quantum theory on classical space-time, and the significance of quantum interference in time} 
The way Level II. is usually constructed, is shown in Fig. 6 below.
\begin{figure}[H]
	\centering
	\includegraphics[width=1.0\linewidth]{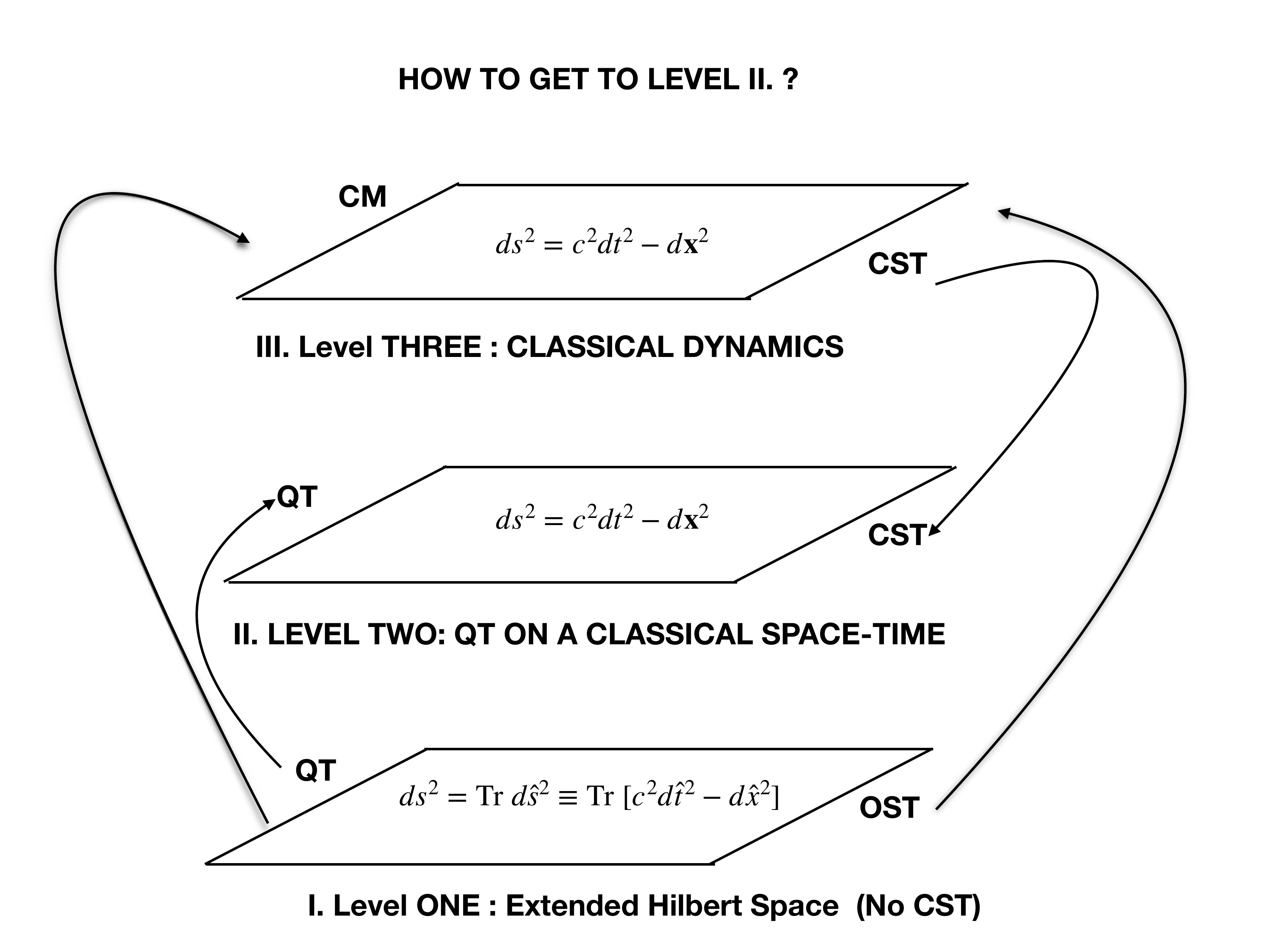}
	\caption{Recovering Level II from Level I.}
	\end{figure}
	That is, we take quantum theory from Level I. (without the postulate of spontaneous localisation) and we take classical space-time from Level III. and we make a hybrid dynamics at Level II. In the light of our discussion in the first section, and in the light of the spontaneous localisation postulate of Level I, we now know that this hybrid dynamics of Level II. cannot be the full story. In fact quantum dynamics can be correctly described only at level I, by using the operator space-time metric. If spontaneous localisation is ignorable (microscopic systems) we get linear quantum theory on an operator space-time, which as we shall soon see, differs from quantum theory on CST by way of predicting interference in time.   If we insist on using a classical space-time, as in Level II., then the minimum we must do is have the GRW theory, expressed by the following standard postulates (non-relativistic theory) \cite{Ghirardi:86,Bassi:03}.
	
1. Given the wave function $\psi ({\bf x_1}, {\bf x_2}, ..., {\bf x_N})$ of an $N$ particle quantum system in Hilbert space, the $n$-th particle undergoes  a `spontaneous collapse' to a random spatial position ${\bf x}$ as defined by the following jump operator:
\begin{eqnarray}
{\psi_{t}({\bf x}_{1}, {\bf x}_{2}, \ldots {\bf x}_{N}) \quad
\longrightarrow \quad} 
 \frac{L_{n}({\bf x}) \psi_{t}({\bf x}_{1},
{\bf x}_{2}, \ldots {\bf x}_{N})}{\|L_{n}({\bf x}) \psi_{t}({\bf
x}_{1}, {\bf x}_{2}, \ldots {\bf x}_{N})\|}
\end{eqnarray}

The jump operator $L_{n}({\bf x})$ is a linear operator defined to be the normalised Gaussian:
\begin{equation}
L_{n}({\bf x}) =
\frac{1}{(\pi r_C^2)^{3/4}} e^{- ({\bf
\hat q}_{n} - {\bf x})^2/2r_C^2}
\end{equation}
${\bf \hat q}_{n}$ is the position operator for the $n$-th particle of the system and the random variable ${\bf x}$ is the spatial position to which the jump occurs. $r_C$, the width of the Gaussian, is a new constant of nature.

The probability density for the $n$-th particle to jump to the position
${\bf x}$ is assumed to be given by:
\begin{equation}
p_{n}({\bf x}) \quad \equiv \quad \|L_{n}({\bf x}) \psi_{t}({\bf
x}_{1}, {\bf x}_{2}, \ldots {\bf x}_{N})\|^2
\end{equation}
Also, it is assumed in the GRW theory that the jumps are distributed in time as
a Poissonian process with frequency $\lambda_{\text{\tiny GRW}}$. This is the second
new parameter of the model.

2. Between
two consecutive jumps, the state vector evolves according to the
standard Schr\"odinger equation.

It is not difficult to see that the GRW theory above can equivalently be expressed by assuming spatial position to be an operator:

We define a set of three new self-adjoint `space operators' ${\bf \hat x}$ which commute with each other and with the ${\bf \hat q}_n$s.. The state of the system is described by the wave function $\psi ({\bf x_1}, {\bf x_2}, ..., {\bf x_N})$, where ${\bf x_n}$ is a set of three degrees of freedom associated with the $n$-th particle, these being real eigenvalues of the newly introduced `space operator' ${\bf \hat x}$ which belongs to the Hilbert space. The state evolves with time according to the following two postulates, which are essentially the same as the GRW postulates, except that one gets rid of classical physical space.

1. Given the wave function $\psi ({\bf x_1}, {\bf x_2}, ..., {\bf x_N})$ of an $N$ particle quantum system in Hilbert space, the $n$-th particle undergoes  a `spontaneous collapse' to a random eigenvalue ${\bf x}$ of ${\bf \hat x}$, as defined by the following jump operator:
\begin{eqnarray}
{\psi_{t}({\bf x}_{1}, {\bf x}_{2}, \ldots {\bf x}_{N}) \quad
\longrightarrow \quad} 
 \frac{L_{n}({\bf x}) \psi_{t}({\bf x}_{1},
{\bf x}_{2}, \ldots {\bf x}_{N})}{\|L_{n}({\bf x}) \psi_{t}({\bf
x}_{1}, {\bf x}_{2}, \ldots {\bf x}_{N})\|}
\end{eqnarray}

The jump operator $L_{n}({\bf x})$ is a linear operator defined to be the normalised Gaussian:
\begin{equation}
L_{n}({\bf x}) =
\frac{1}{(\pi r_C^2)^{3/4}} e^{- ({\bf
\hat q}_{n} - {\bf x})^2/2r_C^2}
\end{equation}
${\bf \hat q}_{n}$ is the position operator for the $n$-th particle of the system and the random variable ${\bf x}$ is the eigenvalue of ${\bf \hat x}$  to which the jump occurs. $r_C$, the width of the Gaussian, is a new constant of nature.

The probability density for the $n$-th particle to jump to the eigenvalue  ${\bf x}$ of ${\bf \hat x}$ is assumed to be given by:
\begin{equation}
p_{n}({\bf x}) \quad \equiv \quad \|L_{n}({\bf x}) \psi_{t}({\bf
x}_{1}, {\bf x}_{2}, \ldots {\bf x}_{N})\|^2
\end{equation}
Also, it is assumed  that the jumps are distributed in time as
a Poissonian process with frequency $\lambda_{\text{\tiny GRW}}$. This is the second
new parameter of the model.

From the structure of these postulates, and from their comparison with the relativistic postulates of Sections II. and III. the following facts are evident: (i) if the operator nature of time is suppressed, and spontaneous localisation is ignored, then relativistic quantum field theory on level I. coincides with relativistic quantum field theory on Level II. (ii) if the operator nature of time is suppressed, and spontaneous localisation is invoked, then one arrives from the relativistic collapse model of Section III. to the non-relativistic GRW theory at Level II. (iii) In order to make a relativistic version of the GRW theory, we must invoke an operator nature for time. 

Thus quantum theory at Level I. differs from quantum theory at Level II, in that at level I. time is an operator, while at level II. it is not. This is the feature that is lost in the hybrid dynamics at level II. What is the evidence for the operator nature of time, and why do we not see it easily? If time is an operator, we should see quantum interference in time. We believe we have a convincing explanation as to why quantum interference in time is so much harder to see than the usual spatial quantum interference. From the relativistic collapse postulates of Section III, and from the GRW postulates, it is plausible to make the assumption that $\eta_{\rm GRW} = \lambda_{\rm GRW}$, and that $ct_C = r_C$. If we assume for $r_C$ the GRW value of $10^{-5}$ cm, then we get that
$t_C=r_C /c\sim 10^{-16}$ s. If we were to make `time slits' with a separation significantly larger than $10^{-16}$ s, then even for microscopic systems, spontaneous collapse in time will destroy quantum interference in time. On the other hand if the time slits have a separation of the order $10^{-16}$ s or smaller, interference in time will be observed. Remarkably enough, attosecond scale  interference in time may have already been observed in the laboratory several years ago \cite{L2005}, and we could possibly consider this to be evidence for the operator nature of time and for the ideas presented in this work. We predict that for time slit separations significantly larger than $10^{-16}$ s, interference in time will not be observed.  Confirmation of this prediction will constitute experimental evidence for relativistic spontaneous localisation in operator space-time, and for collapse of the wave-function as the the mechanism for emergence in space-time.

Outstanding open challenges in this program are generalisation to quantum field theory, and to include gravity. This is currently being attempted. There is perhaps a direct connection of this program, with non-commutative differential geometry.

I would like to thank Angelo Bassi, Kinjalk Lochan, Hendrik Ulbricht,   Bhavya Bhatt, Shounak De, Sandro Donadi,  Priyanka Giri, Anirudh Gundhi, Navya Gupta,  Manish, Ruchira Mishra, Shlok Nahar, Branislav Nikolic, Raj Patil and Anjali Ramesh  for helpful discussions. 

\bibliography{biblioqmtstorsion}

\end{document}